\documentclass[prb,twocolumn,showpacs,preprintnumbers,amsmath,amssymb,superscriptaddress,floatfix]{revtex4}
\usepackage{graphicx}
\usepackage{dcolumn}
\usepackage{bm}
\usepackage{here} 
\usepackage{epsf}
\usepackage{epstopdf}
\usepackage{color}

\begin{document}

\title{Electronic Structures of Ferromagnetic CeAgSb$_2$:\\Soft X-ray Absorption, Magnetic Circular Dichroism\\ and Angle-Resolved Photoemission Spectroscopies}

\author{Yuji Saitoh}
\affiliation{Materials Sciences Research Center, Japan Atomic Energy Agency (JAEA), Sayo, Hyogo 679-5148, Japan}
\author{Hidenori Fujiwara}
\author{Takashi Yamaguchi}
\author{Yasuhiro Nakatani}
\author{Takeo Mori}
\author{Hiroto Fuchimoto}
\author{Takayuki Kiss}
\affiliation{Division of Materials Physics, Graduate School of Engineering Science, Osaka University, Toyonaka, Osaka 560-8531, Japan}
\author{Akira Yasui}
\affiliation{Materials Sciences Research Center, Japan Atomic Energy Agency (JAEA), Sayo, Hyogo 679-5148, Japan}
\affiliation{Japan Synchrotron Radiation Research Institute (JASRI), Sayo, Hyogo 679-5198, Japan}
\author{Jun Miyawaki}
\affiliation{Institute for Solid State Physics (ISSP), University of Tokyo, Kashiwanoha, Chiba 277-8581, Japan}
\affiliation{Synchrotron Radiation Research Organization, University of Tokyo, Sayo-cho, Sayo, Hyogo 679-5198, Japan}
\author{Shin Imada}
\affiliation{Department of Physical Science, Ritsumeikan University, Kusatsu, Shiga 525-8577, Japan}
\author{Hiroshi Yamagami}
\affiliation{Materials Sciences Research Center, Japan Atomic Energy Agency (JAEA), Sayo, Hyogo 679-5148, Japan}
\affiliation{Department of Physics, Faculty of Science, Kyoto Sangyo University, Kyoto 603-8555, Japan}
\author{Takao Ebihara}
\affiliation{Department of Physics, Graduate School of Science, Shizuoka University, Shizuoka 422-8529, Japan}
\author{Akira Sekiyama}
\affiliation{Division of Materials Physics, Graduate School of Engineering Science, Osaka University, Toyonaka, Osaka 560-8531, Japan}

\date{\today}


\begin{abstract}
We report a combined study for the electronic structures of ferromagnetic CeAgSb$_2$ using soft X-ray absorption (XAS), magnetic circular dichroism (XMCD), and angle-resolved photoemission (ARPES) spectroscopies. The Ce $M_{4, 5}$ XAS spectra show very small satellite structures, reflecting a strongly localized character of the Ce $4f$ electrons. The linear dichroism effects in the Ce $M_{4, 5}$ XAS spectra demonstrate the ground state Ce $4f$ symmetry of $\Gamma{_6}$, the spatial distribution of which is directed along the $c$-axis. The XMCD results give support to the picture of local-moment magnetism in CeAgSb$_2$. 
Moreover it is also found that the theoretical band dispersions for LaAgSb$_2$ provides better description of the ARPES band structures than those for CeAgSb$_2$. Nevertheless, ARPES spectra at the Ce $3d$-$4f$ resonance show the momentum dependence for the intensity ratio between Ce $4f^{1}_{5/2}$ and $4f^{1}_{7/2}$ peaks in a part of the Brillouin zone, suggesting the non-negligible momentum dependent hybridization effect between the Ce $4f$ and the conduction electrons. This is associated with the moderate mass enhancement in CeAgSb$_2$.
\end{abstract}

\maketitle

\section{Introduction}

 In $f$-electron materials, strong interactions between $f$- and conduction electrons lead to a rich variety of phenomena~\cite{Stewart1984,Stewart2001,Stewart2006}. Research on such systems has led to a point of view that is summarized by the Doniach phase diagram~\cite{Doniach1977} in which the ground state evolves from a magnetically ordered state with fully localized moments to a paramagnetic heavy-fermion state as a function of tuning parameters, such as pressure, magnetic field, and chemical substitution. At the magnetic/nonmagnetic crossover, the ordering temperature is suppressed to zero and a quantum critical point (QCP) emerges. QCP physics are of great scientific interest since various heavy fermion systems appear to exhibit unconventional superconductivity close to QCP~\cite{Pfleiderer2009}. A competition between local-moment ordering mediated via the Ruderman-Kittel-Kasuya-Yoshida (RKKY) interaction and the onsite Kondo fluctuation of the $f$ electrons is largely responsible for the physics of these materials. The RKKY interaction induces long range magnetic order of the $f$ electrons while the Kondo effect forms magnetic singlet states between the localized $f$ and conduction states. In addition, the crystal-electric-field (CEF) interactions influence the thermodynamic and transport properties~\cite{Levy1989}.
 
\begin{figure}
\begin{center}
\includegraphics[width=7cm,clip]{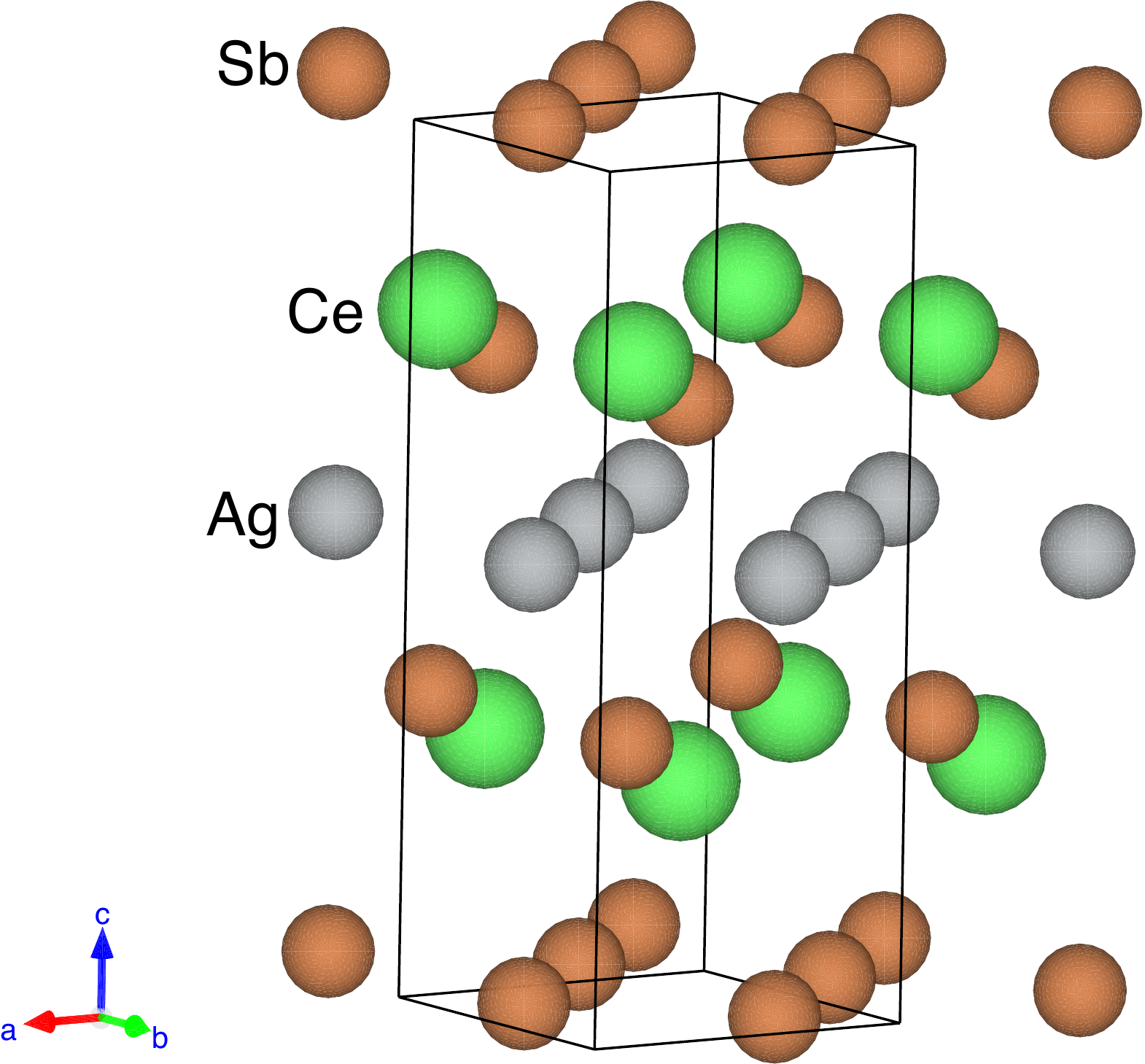}
\caption 
{(color online) Crystal structure of CeAgSb$_2$.}
\label{Fig1}
\end{center}
\end{figure}

It is known that many Ce-based compounds have an antiferromagnetic ground state. CeAgSb$_2$ belongs to a small group of Ce-based ferromagnets with the Curie temperature $T_\textrm{C}$ $\approx$9.7 K at ambient pressure~\cite{Takeuchi2003}. It crystallizes in the tetragonal ZrCuSi$_2$-type structure (space group $P4/nmm$), where planer layers of Sb, Ag and Ce-Sb are stacked along the $c$-axis as shown in Fig.~\ref{Fig1}~\cite{Araki2003,VESTA}. 

Inelastic neutron scattering (INS) experiments and magnetization measurements indicate that the $4f$ electrons are almost completely localized with an ordered magnetic moment of ~0.4 $\mu_{B}$/Ce aligned along the $c$-axis close to that expected for the CEF doublet ground state $\Gamma_{6}$ of Ce$^{3+}$ ions~\cite{Takeuchi2003,Araki2003,Inada2002,Jobiliong2005,Nakashima2003}. The magnetic entropy reaching $\sim$90\% of $R$ln2 at $T_\textrm{C}$, where $R$ is the gas constant, suggests that the Kondo interaction is weaker than the RKKY interaction and the magnetic order develops out of a doublet CEF ground state~\cite{Takeuchi2003,Araki2003,Inada2002}. The localized $4f$ picture is also consistent with the de Haas-van Alphen (dHvA) experiments on CeAgSb$_2$~\cite{Inada2002}. Many ferromagnetic Ce compounds are considered to be well in the localized moment regime in which they do not possess enhanced effective masses, which contrasts with the typical heavy fermions.

Although the application of pressure toward a critical pressure $P_c\approx$3.3 GPa suppresses $T_\textrm{C}$, no superconductivity is found above 90 mK~\cite{Nakashima2003}. To the best of our knowledge, there are no ferromagnetic Ce systems that show superconductivity. This contrasts with the superconductivity in ferromagnetic uranium compounds such as UGe$_2$, URhGe and UIr~\cite{Pfleiderer2009}. 
 
A moderately enhanced Sommerfeld coefficient $\gamma$ = 46 mJ/mol$\cdot$K$^2$ corresponds to a mass enhancement of $\approx$21 compared to the nonmagnetic La analog~\cite{Inada2002}. The characteristic behavior of resistivity in the paramagnetic phase suggests the existence of the Kondo effect in CeAgSb$_2$. A relatively high Kondo temperature $T_\textrm{K}\sim$23 K, which is comparable to that for a typical heavy fermion compound CeRu$_2$Si$_2$ ($\gamma$ = 350 mJ/mol$\cdot$K$^2$) with highly delocalized $4f$ electrons~\cite{Aoki2014}, has been reported for the CEF ground state doublet~\cite{Jobiliong2005}. Previous angle-resolved photoemission (ARPES) for CeAgSb$_2$ indicates that the experimental band structures and the topology of the Fermi surfaces near the $\Gamma$ point in the Brillouin zone (BZ) are similar to those for LaAgSb$_2$~\cite{Arakane2007}. However, the excitation photon energy ($h\nu$) was fixed to be 21.2 eV, and thus the band structures were probed in the limited region of the BZ. Since the band structures predicted by a fully relativistic linearized augmented-plane-wave (FLAPW) method for CeAgSb$_2$ and LaAgSb$_2$~\cite{Inada2002} show the three-dimensional character though the Fermi surfaces are highly two-dimensional as reported by the dHvA experiments, it is important to reveal the band dispersion in the three-dimensional Brillouin Zone. Moreover, it is well known that the conventional low energy ARPES probes only the surface electronic structures which is often quite different from the bulk ones especially for the strongly correlated $4f$ electron systems~\cite{Sekiyama2000a,Sekiyama2000b,Iwasaki2002}. In contrast, the soft X-ray ARPES is more sensitive to the bulk electronic structure with still reasonable momentum resolution~\cite{Yano2007,Yano2008,Fujiwara2015a}.

In this work we show a combined spectroscopic study with soft X-ray absorption (XAS), magnetic circular dichroism (XMCD), and ARPES for revealing the relation between the magnetism and the hybridization between the Ce $4f$ and the conduction electrons (c-$f$ hybridization) for CeAgSb$_2$.

\section{Experiment}

High-quality CeAgSb$_2$ single crystal samples were grown by the Sb-self-flux method~\cite{Takeuchi2003}. The spectroscopic measurements XAS, XMCD and ARPES were performed at the twin-helical undulator beamline BL23SU of SPring-8~\cite{Saitoh2012}. The single crystal samples were cleaved under the base pressure of about 1.1$\times$10$^{-8}$ Pa to expose clean (001) surfaces. XAS and XMCD spectra were obtained in total electron yield mode with an energy resolution of 80 meV, using a superconducting magnet with fields up to $\pm$10 T along the incident beam direction. The XAS spectra without a magnetic field were measured in on-the-fly scanning mode, while the XAS and XMCD spectra under magnetic fields were measured in 1 Hz helicity-switching mode~\cite{Saitoh2012}. In order to eliminate any experimental artifacts arising from system errors, each XMCD spectrum was measured for opposite orientations of the applied magnetic field and the resulting spectra were averaged. The incident photon energy was calibrated to $h\nu$ = 867.12 eV for the $1s$-$3p$ transition peak of Ne gas. ARPES spectra were acquired using a Gammadata-Scienta SES-2002 electron analyzer. The energy resolution was set to 100 meV for Ce $3d$-$4f$ resonant ARPES experiments, while about 240 meV for the off-resonant ARPES. The momentum resolution was set to $\pm0.1^\circ$ ($0.15^\circ$) parallel (perpendicular) to the analyzer slit. The sample temperature was set to 20 K for the ARPES measurements.

\section{Results and Discussion}

\subsection{Soft X-ray Absorption Spectroscopy}

\begin{figure}
\begin{center}
\includegraphics[width=7cm,clip]{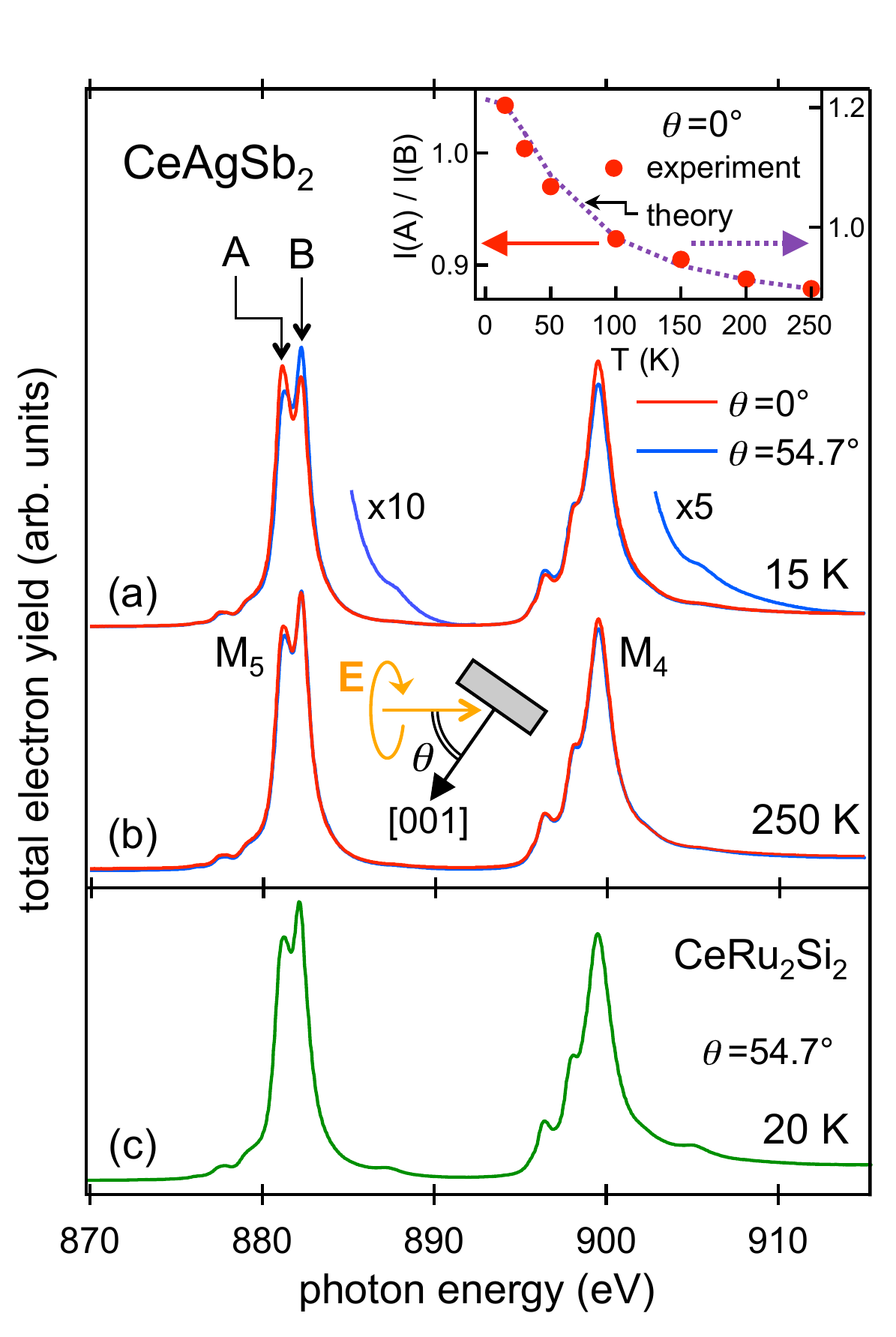}
\caption 
{(color online) Ce $M_{4,5}$-edge circularly polarized XAS spectra of CeAgSb$_2$ for x-ray incidence angles of $\theta= 0^\circ$ and 54.7$^\circ$ with respect to the crystalline $c$-axis parallel to the surface normal measured at 15 K (a) and 250 K (b) in the paramagnetic phase without applied magnetic field. Top inset: Temperature dependence of the peak A to peak B intensity ratio for $\theta= 0^\circ$ spectra (circles), together with the simulation results (dotted line) based on the multiplet calculation as discussed in the text. Bottom inset: Schematically drawn experimental geometry. (c) Ce $M_{4,5}$-edge circularly polarized XAS spectra of CeRu$_2$Si$_2$ recorded at 20 K with the $\theta= 54.7^\circ$ geometry.}
\label{Fig2}
\end{center}
\end{figure}

Figure~\ref{Fig2} shows the circularly polarized Ce $M_{4,5}$ XAS spectra of CeAgSb$_2$ and CeRu$_2$Si$_2$ taken in the paramagnetic phase at zero applied magnetic field. In Fig.~\ref{Fig2}(a), the XAS spectra of CeAgSb$_2$ measured at 15 K for photon incidence angles of $\theta=0^\circ$ and 54.7$^\circ$ relative to the $c$-axis (surface normal) are shown. A clear difference in the spectral line shape is observed. In particular, the two prominent multiplet peaks at the Ce $M_{5}$ edge, labeled A and B, show that the peak A is larger for $\theta$ = 0$^\circ$ and the peak B is larger for 54.7$^\circ$. The XAS lineshape at $\theta=0^\circ$ shows pronounced temperature dependence, resulting in much reduced anisotropy at 250 K as shown Fig.~\ref{Fig2}(b). The upper inset in Fig.~\ref{Fig2} shows the intensity ratio of the peaks A and B, $I(A)/I(B)$, measured from the pre-edge background for $\theta=0^\circ$ plotted as a function of temperature (full circles) for the scale on the left axis. It has been known that XAS spectra exhibit orientation dependence on non-cubic materials, even for an unpolarized beam~\cite{Brouder1990}, since the electric field vector $\vec{E}$ must be transverse to the direction of light propagation. In the absence of a magnetic field, the angular dependence of the dipole-allowed absorption for circular polarized light is the same as that for unpolarized light~\cite{Brouder1990}. The magic angle condition of $\theta=54.7^\circ$ provides a polarization-averaged isotropic XAS spectrum~\cite{Stoehr1995} $\mu_{\textrm{iso}}=(2\mu_{\perp{c}}+\mu_{{\parallel}{c}})/3$, where $\mu_{\perp{c}}$ ($\mu_{{\parallel}{c}}$) is the XAS intensity with linear polarization $\vec{E}$ perpendicular (parallel) to the tetragonal $c$-axis. This is because the angular dependence of the XAS is expressed as $\mu_{\theta}=\mu{_0}{\cos}{^2}{\theta}+\mu{_{90}}{\sin}{^2}{\theta}$, where $\mu_{0}=\mu_{\perp{c}}$ and $\mu_{90}= (\mu_{\perp{c}}+\mu_{{\parallel}{c}})/2$, so that the isotropic spectrum is obtained from $3\cos{^2}{\theta}=1$~\cite{Stoehr1995}. 

The shape of the XAS spectra at $\theta=54.7^\circ$ is characteristic of a localized electronic configuration close to $4f^1$ (Ce$^{3+}$), similar to those for polycrystalline $\gamma$-like Ce compounds~\cite{Kaindl1985}. The very weak satellite structures observed at the high energy tail above each edge 
indicate the presence of the finite c-$f$ hybridization. The intensity of the satellite relative to the main peak is a reliable qualitative guide to the initial state $4f^0$ component $1-n_f$, where $n_f$ is the Ce $4f$ electron occupation number~\cite{Fuggle1983}. The comparison of the isotropic XAS spectra in Fig.~\ref{Fig2}(a) and \ref{Fig2}(c) illustrates that the c-$f$ hybridization is much smaller for CeAgSb$_2$ than for CeRu$_2$Si$_2$.

The observed XAS anisotropy arises from the difference between $\mu_{\perp{c}}$ and $\mu_{{\parallel}{c}}$ (namely, linear dichcroism, LD) and is attributable to the anisotropy of the initial state $4f$ charge distribution induced by CEF effects~\cite{Hansmann2008}. The tetragonal crystal field acting on a Ce$^{3+}$ ion splits the $J = 5/2$ ground state multiplet into three KramerÕs doublets~\cite{Aviani2001} as, 
\begin{eqnarray}
\Gamma{_7}{^1} & = & \alpha \left | \pm 5/2 \right > + \beta \left | \mp 3/2 \right >,\\
\Gamma{_7}{^2} & = & - \beta \left | \pm 5/2 \right > + \alpha \left | \mp 3/2 \right >,\\
\Gamma{_6} & = & \left | \pm1/2 \right>, 
\end{eqnarray}
with $\alpha^2+\beta^2=1$ ($\alpha, \beta>0$). It has been shown that the analysis of the LD effects using the ligand field multiplet (LFM) model gives a good quantitative description of the CEF ground state~\cite{Hansmann2008,Willers2010,Willers2012,Willers2014}. In CeAgSb$_2$, the CEF splitting energies from the ground state to the first and second excited states ($\Delta_1$ and $\Delta_2$) have been determined to be 5.2 meV ($\approx$60 K) and 12.5 meV ($\approx$145 K) from the INS measurements, in which the CEF ground state is $\left | \pm1/2 \right >$ and the first and second exited levels are dominated by $\left | \pm3/2 \right >$ and $\left | \pm 5/2 \right >$, respectively~\cite{Araki2003}. This CEF scheme is consistent with thermodynamic experiments~\cite{Takeuchi2003,Jobiliong2005}.

\begin{figure}
\begin{center}
\includegraphics[width=7cm,clip]{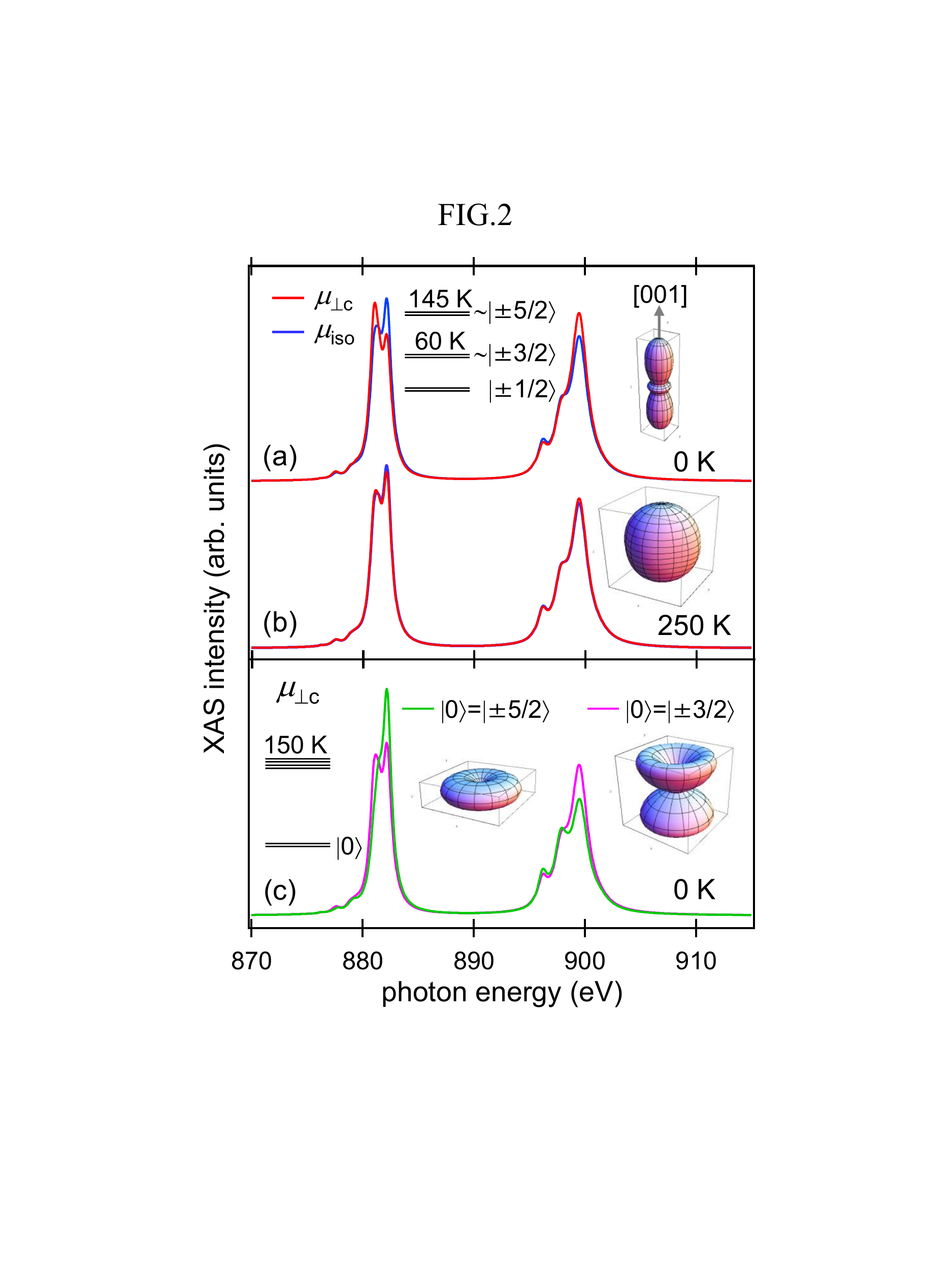}
\caption 
{(color online) Ligand field multiplet calculations for the Ce$^{3+}$ $M_{4,5}$ XAS $\mu_\textrm{iso}$ and $\mu_{\perp{c}}$ spectra at 0 K (a) and 250 K (b) using a tetragonal CEF scheme of CeAgSb$_2$, as depicted by the upper inset, determined by neutron scattering experiments. (c) The $\mu_{\perp{c}}$ spectra calculated for the $\left | \pm 3/2 \right >$ and $\left | \pm 5/2 \right >$ ground states with a CEF splitting of 150 K (c). Also shown are the corresponding initial-state $4f$ charge distributions.}
\label{Fig3}
\end{center}
\end{figure}

Figures~\ref{Fig3}(a) and~\ref{Fig3}(b) show the results of multiplet calculations for Ce$^{3+}$ $3d{^{10}}4f{^1}\to3d{^9}4f{^2}$ electric-dipole allowed transition using the XTLS 8.3 program~\cite{Tanaka1994}. The calculational method follows that of Ref.~\onlinecite{Hansmann2008}, which takes into account the intra-atomic $4f$-$4f$ and $3d$-$4f$ Coulomb interactions, the atomic $3d$ and $4f$ spin-orbit couplings, and local crystal field interactions. On the basis of the well-established Hartree-Fock values for Ce$^{3+}$~\cite{Thole1985}, the crystal field parameters obtained from the INS measurements~\cite{Araki2003} are the fundamentally relevant quantities. In Fig.~\ref{Fig3}(a), the calculated $\mu_\textrm{iso}$ and $\mu_{\perp{c}}$ spectra at 0 K are displayed. For comparison, $\mu_{\perp{c}}$ spectra calculated for the $\left | \pm 3/2 \right >$ and $\left | \pm 5/2 \right >$ ground states with ${\Delta_1}={\Delta_2}=150$ K are shown in Fig.~\ref{Fig3}(c). These results immediately give a direct verification that the wave function of the ground state is $\left | \pm 1/2 \right >$ of the $\Gamma_6$ symmetry, because the $\mu_{\perp{c}}$ spectrum of a $\Gamma_7$ ground state is given by a linear combination of those for the $\left | \pm 3/2 \right >$ and $\left | \pm 5/2 \right >$ states with $I(B)$ larger than $I(A)$. In addition, as shown in the inset of Fig.~\ref{Fig2} and Fig.~\ref{Fig3}(b), the qualitative behavior of the experimentally observed temperature dependence is accounted for by the calculations, the initial state of which is weighted with the Boltzman distribution. The decrease in the observed anisotropy with increasing temperature is caused by the progressive thermal population of the excited CEF states, thereby reducing the anisotropy of the $4f$ wave functions as shown insets in Fig.~\ref{Fig3}.

\begin{figure}
\begin{center}
\includegraphics[width=7cm,clip]{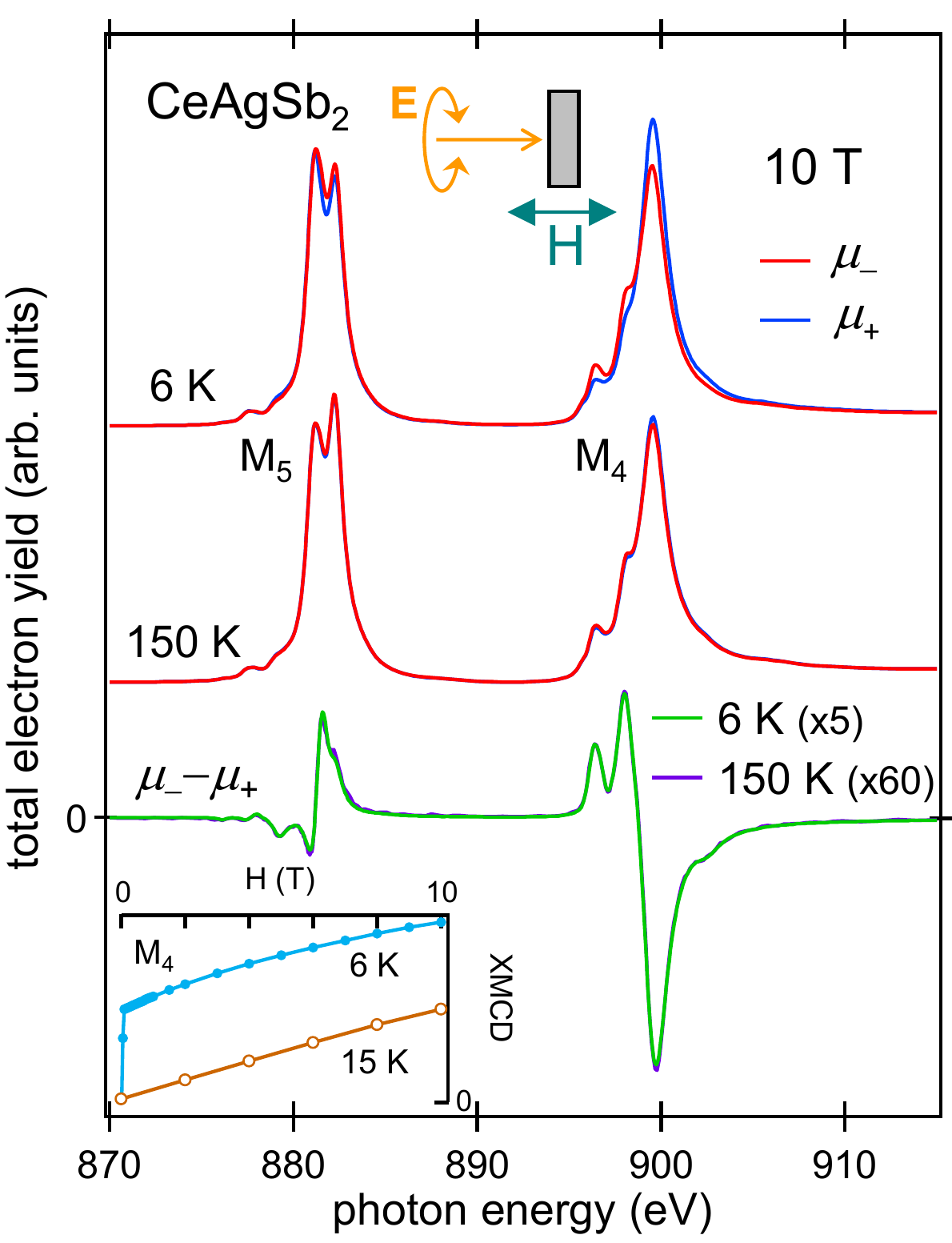}
\caption 
{(color online) Ce $M_{4,5}$ XAS and XMCD spectra of CeAgSb$_2$ for parallel ($\mu_{+}$) and antiparallel ($\mu_{-}$) alignment of photon helicity and sample magnetization directions for $\theta= 0^\circ$ at 6 and 150K. The lower inset shows the XMCD magnetization curves for 6 and 15 K at the $M_4$ edge.}
\label{Fig4}
\end{center}
\end{figure}

Figure~\ref{Fig4} presents the Ce $M_{4,5}$ XAS ($\mu_{+}$ and $\mu_{-}$) and XMCD (${\mu_{-}}-{\mu_{+}}$) spectra of CeAgSb$_2$ measured at 6 K and 150 K in an external field of $H = 10$ T for $\theta= 0^\circ$. Here, $\mu_{+}$ and $\mu_{-}$ refer to the XAS intensity recorded for parallel and antiparallel alignment of photon helicity and sample magnetization. The inset at lower left displays the element-specific magnetization curves at 6 and 15 K recorded at the $M_{4}$ edge (899.6 eV) while sweeping the applied magnetic field, showing ferromagnetic and paramagnetic responses, respectively. In contrast to the $\mu_{\perp{c}}$ curves in Fig.~\ref{Fig2}, the spectral XMCD shape remains the same with increasing temperature. The multiplet features of the XMCD line shape are characteristic of $\gamma$-like Ce compounds~\cite{Schille1994,Ishikawa2004,Okane2012}.

\begin{figure}
\begin{center}
\includegraphics[width=7.3cm,clip]{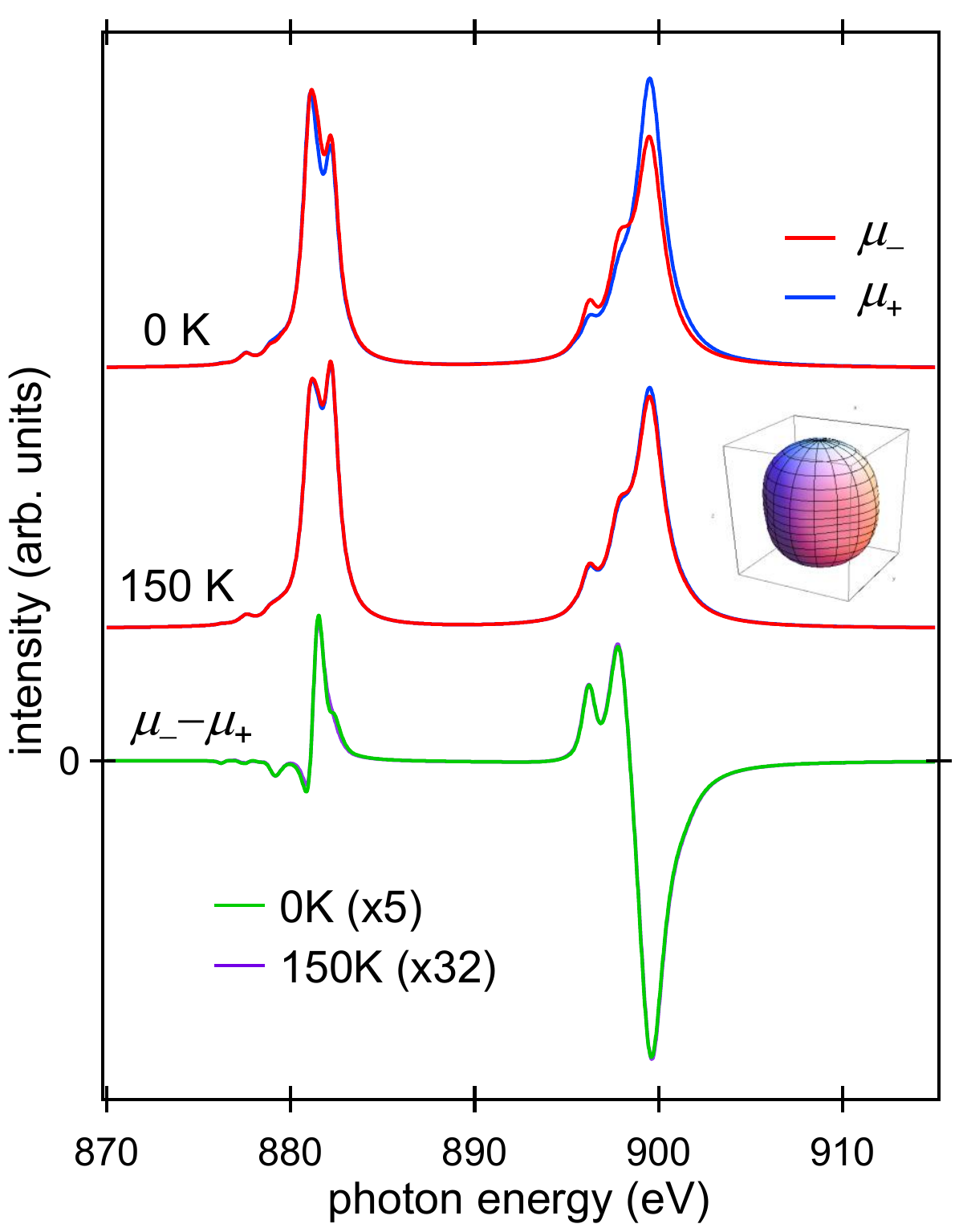}
\caption 
{(color online) Ligand field multiplet calculations for the Ce $M_{4,5}$ XAS and XMCD spectra of CeAgSb$_2$ together with the initial-state $4f$ charge distribution at 150 K. }
\label{Fig5}
\end{center}
\end{figure}

Figure~\ref{Fig5} shows the calculated XAS and XMCD spectra for 0 K with an infinitesimal exchange field and for 150 K with an external magnetic field of 10 T, which are in good agreement with our experimental spectra. The constant XMCD shape is associated with a constant ratio of the orbital to spin magnetic moment $m{_L}/m{_S}$ of $-4$ for Ce$^{3+}$ states that split off from a pure $J=5/2$ state by CEF~\cite{Schille1994}. Quantitative information on the total atomic magnetic moment $m_{total}=m{_S}+m{_L}$ of a Ce site separated into $m_S = -2\left< {S_z} \right>\mu_B$ and $m_L=-\left< {L_z} \right>\mu_B$ can be obtained by applying the corrected XMCD sum rules~\cite{Thole1992,Carra1993,Teramura1996},
\begin{eqnarray}
\left< {L_z}\right> & = & \frac{2q(14-{n_f})}{r{P_c}},\\
\frac{\left< {L_z}\right>}{\left< {S_z}\right>} & = & \frac{4C}{5{\frac{p}{q}}-3} \left ( 1+3\frac{\left< {T_z} \right>}{\left< {S_z} \right>}\right ), 
\end{eqnarray}
where $p$ ($q$) is the integral of the XMCD signal over the $M_5$ edge ($M_{4,5}$ edges), $r$ is the integral of the polarization-summed XAS intensity, ${\mu_{+}}+{\mu_{-}}$, over the $M_{4,5}$ edges~\cite{comments_LD}, $P_c$ is the circular polarization rate of the incident X-rays and $C$ (=1.60 for a Ce$^{3+}$ ion~\cite{Teramura1996}) is the correction factor for the mixing of the multiplet structure between the $3d_{5/2}$ and $3d_{3/2}$ levels caused by $3d$-$4f$ electrostatic interactions. $\left< {T_z} \right>$ is the expectation value of the magnetic dipole operator. Assuming $P_{c}=0.97$~\cite{Saitoh2012,Hirono2005}, $n_{f}=1$ and an atomic $\left< {T_z} \right> / \left< {S_z} \right>$ ratio of 8/5~\cite{Teramura1996}, we obtained $m_{total}=0.35\pm0.02$ $\mu_{B}$/Ce and $m{_L} / m_{S} = -4.05 \pm 0.05$ from the XAS and XMCD spectra at 6 K. The cutoff energy of $p$ was set to the minimum of the polarization-summed XAS spectrum between the $M_5$ and $M_4$ edges and that of $q$ to 940 eV. This $m_{total}$ value is close to the saturation moment of $\sim0.4\mu_B$/Ce at 1.5 K determined from bulk magnetometry~\cite{Takeuchi2003,Nakashima2003,Araki2003,Inada2002,Jobiliong2005}. This provides additional evidence that the localized $4f$ electrons in the CEF dominate the magnetic properties of CeAgSb$_2$. 

The $m_{total}$ at 150K is estimated to be approximately 0.03$\mu_{B}$/Ce, while the theoretical XMCD signal arises from $m_{total}=0.067\mu_{B}$/Ce. This theoretical value is close to those expected from the magnetic susceptibility data~\cite{Takeuchi2003,Araki2003,Inada2002,Jobiliong2005}. The smaller value of the experimental $m_{total}$ is attributable to some possible effects: the magnetic contribution of the non-$4f$ electrons as predicted by theory~\cite{Inada2002} and the surface effects including so-called saturation effects in the total electron yield measurement~\cite{Nakajima1999}. Additional insight concerning the magnetism in CeAgSb$_2$ would be provided through the further use of element- and orbital-specific XMCD.

\subsection{Soft X-ray Angle-Resolved Photoemission Spectroscopy}

\begin{figure}
\begin{center}
\includegraphics[width=6.8cm,clip]{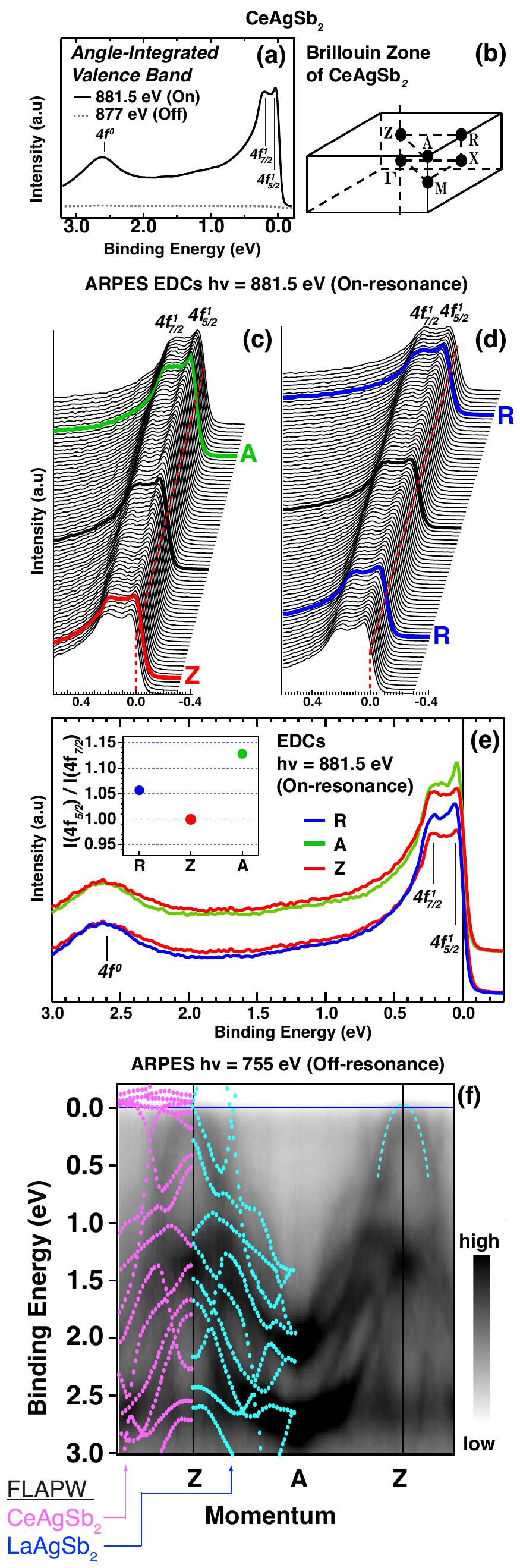}
\caption 
{(color online) (a) Angle-integrated valence band photoemission spectra for CeAgSb$_2$ measured at $h\nu=877$ eV(dashed line) and 881.5 eV (solid line) with off- and on-resonance conditions, respectively. (b) Brillouin zone for CeAgSb$_2$. (c) EDCs obtained along the Z-A cut, and (d) those for R-R lines. The thin doted lines are guide to the eye indicating the $E_\textrm{F}$. (e) Selected EDCs recorded at the high symmetry points by the Ce $3d$-$4f$ resonant ARPES. (f) The intensity plot of ARPES spectra along the Z-A lines at  $h\nu=755$ eV in an off-resonant condition. The dashed curve indicates the hole-like band centered around the Z point. The FLAPW band structures along the Z-A cut~\cite{Inada2002} (dots) for CeAgSb$_2$(left) and LaAgSb$_2$ (right) are superimposed on (f).}
\label{Fig6}
\end{center}
\end{figure}

The $4f$ electronic structure of CeAgSb$_2$ can be directly observed using a Ce $3d$-$4f$ resonant photoemission technique~\cite{Sekiyama2000a,Sekiyama2000b,Iwasaki2002} through the drastic enhancement of the Ce $4f$ weights as shown in Fig.~\ref{Fig6}(a). In the single impurity model picture~\cite{Malterre1996}, the three-peaked structure at resonance is associated with $4f^{1}_{5/2}$, $4f^{1}_{7/2}$ and $4f^{0}$ final states. The spectral weight of the $4f^{1}_{7/2}$ peak around 0.2 eV is almost the same as that of the $4f^{1}_{5/2}$ peak in the vicinity of the Fermi level ($E_\textrm{F}$), which is typical of $\gamma$-like low $T_\textrm{K}$ Ce compounds~\cite{Suga2002}. This is consistent with the small satellite structures observed in Fig.~\ref{Fig2}, again indicative of the weak c-$f$ hybridization effects, which allow the moderate mass enhancement below the magnetic ordering temperature in CeAgSb$_2$. We have further studied the Ce $3d$-$4f$ resonant ARPES to reveal momentum dependent effects in the Ce $4f$ states. Note that at the on-resonant condition with $h\nu$ = 881.5 eV (at the peak A in Fig.~\ref{Fig2}) one can probe the electronic structures in the Z-A-R plane of the three dimensional BZ shown in Fig.~\ref{Fig6}(b)~\cite{innerPotential}. In the energy distribution curves (EDCs) along the Z-A and R-R lines in Fig.~\ref{Fig6}(c) and (d), one can directly find the non-dispersive structures for both $4f^{1}_{7/2}$ and $4f^{1}_{5/2}$ components, showing the localized nature of the Ce $4f$ states. The relative intensity of the two peaks, however, gradually changes upon sweeping from the Z to A points as shown in Fig.~\ref{Fig6}(c), yet it is not the case along the R-R cuts in Fig.~\ref{Fig6}(d). These features are highlighted by the comparison of the selected EDCs recorded at the Z, A and R points in Fig.~\ref{Fig6}(e), where the spectra are normalized by the intensity of the $4f^{0}$ satellite. The intensity ratio of the $4f^{1}_{5/2}$ to $4f^{1}_{7/2}$ peaks is plotted in the inset of Fig.~\ref{Fig6}(e). Although the intensity modulation observed in CeAgSb$_2$ is much smaller than that observed in more strongly hybridized heavy fermion systems~\cite{Denlinger2001,Danzenbaecher2005,Im2008}, it can be related to the Kondo lattice effect, providing detailed information about the weak c-$f$ hybridization effects. 

Figure~\ref{Fig6}(f) shows the conduction band structure obtained along the Z-A line at $h\nu$ = 755 eV with an off-resonant condition. In contrast to the no appreciable dispersive features in Ce $4f$ resonant ARPES, the dispersive bands due to the non-$4f$ electrons are clearly observed. We have compared the results with published FLAPW band structures for CeAgSb$_2$ and LaAgSb$_2$~\cite{Inada2002}: the latter corresponds to the reference of the localized limit for the $4f$ electrons. The superimposed FLAPW calculations for both CeAgSb$_2$ and LaAgSb$_2$ well capture the general features of the experimental band dispersions. However, the hole-like dispersion centered around the Z points ranging from 0.5 eV to $E_\textrm{F}$ in the ARPES is better reproduced by the FLAPW bands for LaAgSb$_2$ supporting the localized picture of the Ce $4f$ electrons. In addition, the overall band structures for LaAgSb$_2$ probed by soft X-ray ARPES measurements are quite similar to those for CeAgSb$_2$ (not shown)~\cite{Fujiwara2015b}. 

We should note that the spatial distribution of the ground state Ce $4f$ wave function with $\left | \pm 1/2 \right >$ symmetry points along the $c$-axis as shown in Fig.~\ref{Fig3} and each Ce atom has no neighboring atoms along the $c$-axis as shown in Fig.~\ref{Fig1}. The essentially localized nature of the $4f$ electrons in CeAgSb$_2$ can be understood as a natural consequence of the local atomic environment. On the other hand, this anisotropic spatial distribution is attributable to the small but observable momentum dependent c-$f$ hybridization effects.

Finally it should be pointed out that the momentum dependence of the $4f$ features shown in Figs.~\ref{Fig6}(c)-(d) cannot be accommodated by a simplified periodic Anderson model~\cite{Denlinger2001,Danzenbaecher2005}, since the conduction bands have a bottom at the A point and crosses the Ce $4f$ states around the Z point as shown Fig.~\ref{Fig6}(f). This suggests that in extracting detailed information from the experiment the momentum dependence of the c-$f$ hybridization strength should be taken into account in realistic effective models with the Ce $4f$ ground state of $\Gamma{_6}$ symmetry.

\section{CONCLUSION}

We have performed a combined soft X-ray spectroscopy for CeAgSb$_2$. The results indicate that CeAgSb$_2$ can be classified as a nearly local-moment system in which the Kondo screening is week and the RKKY interaction dominates to form a ferromagnetic ground state. Except for the small satellite structure, our experimental XAS and XMCD data are well reproduced by the calculations in the framework of the LFM model, demonstrating the CEF scheme with a $\Gamma_{6} =\left | \pm 1/2 \right >$ ground state without ambiguity. The localized $4f$ picture is corroborated by the non-dispersive Ce $4f$ states probed by the Ce $3d$-$4f$ resonant ARPES and the significant similarity in the dispersive band structures between the experimental results of CeAgSb$_2$ and the FLAPW calculations for LaAgSb$_2$. In addition, small c-$f$ hybridization effects coherently coupled with the lattice are observed in the Ce $4f^{1}_{5/2}$ peak intensity of the resonant ARPES spectra. This allows the moderate mass enhancement in CeAgSb$_2$.

\section*{Acknowledgments}
We acknowledge the help and support of Y. Takeda of JAEA during the beamtime. We also thank Y. Nakata, S. Naimen, K. Takeuchi and T. Yagi for supporting the measurements. The measurements were performed under the approval of BL23SU at SPring-8 (Proposal No. 2011B3834, 2012A3834 and 2012B3834). This work was financially supported by a Grants-in-Aid for Scientific Research on Innovative Areas (JP20102003 and JP16H01074), Grants-in-Aid for Young Scientists (JP23740240), and that for Scientific Research (JP16H04014) from the Ministry of Education, Culture, Sports, Science and Technology, Japan.


\begin{thebibliography}{9}
\bibitem{comments_LD}
Here we have neglected the LD effects with use of generally assumed relation $\mu_\textrm{iso} = (\mu_{+} + \mu_{-})/2$. The LFM calculations shown in Fig.~\ref{Fig5} indicate that this results in an underestimation of $\left<L{_z}\right>$ by less than 3\%.
\bibitem{innerPotential}
The inner potential for CeAgSb$_2$ is evaluated as 4 eV on the basis of the periodicity of the band dispersions along the $\Gamma$-Z direction.
\end{thebibliography}
\end{document}